\documentclass[pre,aps,showpacs,twocolumn,floatfix,superscriptaddress]{revtex4}

\usepackage{graphicx}
\usepackage{bm} 
\usepackage{epsfig}
\usepackage{xcolor}
\usepackage{enumitem}
\usepackage[hidelinks]{hyperref}
\usepackage{amsmath}
\usepackage{float}%
\usepackage{hyperref}
\usepackage{siunitx}
\usepackage{booktabs}

\setlist[itemize]{leftmargin=*}
\setlist[enumerate]{leftmargin=*}	
\newcommand{\be}{\begin{equation}} 
\newcommand{\ee}{\end{equation}}
\newcommand{\bea}{\begin{eqnarray}}
\newcommand{\eea}{\end{eqnarray}}
\newcommand{\ba}{\begin{array}}
\newcommand{\ea}{\end{array}}

\begin{document}

\title{Hippocampal synchronization in a realistic CA1 neuron model.}

\author{Alessandro Fiasconaro}
\email{afiascon@unizar.es}
\affiliation{Dpto. de F\'isica de la Materia Condensada,
Universidad de Zaragoza. 50009 Zaragoza, Spain}
\affiliation{Instituto de Biocomputaci\'on y F\'isica de Sistemas
Complejos, Universidad de Zaragoza. 50018 Zaragoza, Spain}
\affiliation{Istituto di Biofisica, Consiglio Nazionale delle Ricerche. Palermo, Italy}

\author{Michele Migliore}
\email{migliore@cnr.it}
\affiliation{Istituto di Biofisica, Consiglio Nazionale delle Ricerche. Palermo, Italy}

\date{\today}

\begin{abstract}
This work delves into studying the synchronization in two realistic neuron models using Hodgkin-Huxley dynamics. Unlike simplistic point-like models, excitatory synapses are here randomly distributed along the dendrites, introducing strong stochastic contributions into their signal propagation. To focus on the role of different excitatory positions, we use two copies of the same neuron whose synapses are located at different distances from the soma and are exposed to identical Poissonian distributed current pulses. The synchronization is investigated through a specifically defined spiking correlation function, and its behavior is analyzed as a  function of several parameters: inhibition weight, distance from the soma of one synaptic group, excitatory inactivation delay, and weight of the excitatory synapses.
\end{abstract}

\keywords{Stochastic Modeling, Fluctuation phenomena, Single neuron modeling, Neural systems}

\maketitle

\section{Introduction.}
As a part of the limbic system, the mammalian hippocampus, located in the allocortex brain region, is responsible of both short-term and long-term memory, and consequently plays an important role in spatial navigation. It is primarily affected in Alzheimer's disease, so, neuron activity in the hippocampal region is crucial for cognitive brain functions~{\color{violet}\cite{1993Okeefe}}. 

The CA1 pyramidal neurons of the hippocampus receive signals from the entorhinal cortex. Specifically, they receive two different inputs: one directly from the entorhinal cortex (known as the {\it direct path}), the other from the CA3 pyramidal neurons. The latter receive the signal from the granule cells, which in turn receive the signals from the entorhinal cortex (completing the so-called {\it trisynaptic pathway}, so called because three sets of synapses are needed to connect the neurons that propagate along this pathway to the CA1 neurons). 

The fundamental mechanism of spiking is that the post-synaptic signals generated in the dendrites are able to reach the soma both almost simultaneously and with appropriate intensity. In other words, the signals must be integrated in the soma within a suitable time window to produce the significant output represented by the depolarisation of the soma.

Synchronization of neuronal activity is then crucial for the proper functioning of brain networks, and the coordinated timing of spiking activity emerges as a compelling candidate mechanism for encoding information, recognition and cognition~\cite{2015Fries,1993Okeefe,1995Mainen, 1995Lisnan,1995Hopfield,2000HopfieldBrody}.

Numerous papers have studied network synchronization in the framework of different models such as Kuramoto oscillators~\cite{1975Kuramoto,2005Ritort}, Izhikewich model~\cite{2003Izhikewich}, integrated and fire (LIF) schemes~\cite{1999Abbott-LIF} along with their various generalizations~\cite{2005Brette,2021Gorski,2023Marasco}, or by using the complete Hodgkin-Huxley set of equations~\cite{1952HH}, as in the case faced here.

Despite the many fundamental insights into synchronization phenomena that these works have satisfactorily provided, several concerns limit the universal application of these results in understanding functional brain synchronization mechanisms: The main limitations in many of the above mentioned models are that either the neurons are modeled as single dynamical entities without internal structure, or the dynamics lacks the inclusion of inhibition mechanisms as in the Kuramoto-type models.
Point-like neurons have no internal delay due to signal propagation, which can lead to energy loss in their dynamics and even failure of the spiking process.
Moreover, many phenomenological mesoscopic models do not clarify the relationship between their parameters and the biological quantities involved in neuronal function. 

In this sense, an overly coarse-grained model reduces the ability to control biological parameters (such as excitatory or inhibitory synaptic conductances, rise/decay times in gate currents, and their intensities) and risks impeding the biochemical understanding of the processes. This, in turn, diminishes the ability to propose or test parameter-oriented therapeutic approaches.

For this reason, more realistic advances in neuronal modeling have gradually been developed by incorporating the contributions of dendritic extension~\cite{1999Magee,1998Crook} and more detailed morphological models of many types of neurons~\cite{2008Gunay,2018PLoS_migliore}.

Regarding the spatial extension, real neurons exhibit a stochastic distribution of excitatory synapses, which complicates the electrodynamic efficiency of current propagation along the dendrites also because of their geometry~\cite{2005Stuart,2015LiGirault} or the detailed localization of their different ion channels~\cite{2002Migliore}. In other words, the different geometric structure of neurons and their synaptic distribution generate different signal phases upon arrival at somas~\cite{2010Schultheiss}.

The axons originating from the CA3 neurons of the {\it trisynaptic path} transmit their signals to the apical dendrites of the CA1 neurons. The positioning of CA1 excitatory synapses varies due to slight differences in the axon trajectories reaching different neurons and/or the different heights of neurons within the brain network layer. As a result, the relative distance of these signals from the soma varies between different CA1 neurons.

This study focuses on this specific variability examining the CA1 spiking as the result of the synchronization of the signals arriving at the apical dendrites from the same pathway that originates in the CA3 neurons, by taking into account their different spatial distributions, and verifying the robustness of this synchronization over changes in biological parameters such as synapse distance from the soma, inhibitory interneuron weights, rates of the arriving currents, and time inactivation delay of the excitatory synapses.

For this purpose, we use two identical neurons whose synapses are activated by identical signals. We analyze the synchronization of the spikes using a newly developed correlation measure based on the concept of the difference of their phases, the latter being defined for each neuron at any time and normalized with respect to the time lapse between successive spikes.

The paper is organized as follows: In Sec.~\ref{methods} we provide an explanation of the model, along with the procedure used in the simulations and the definition of the correlation measure to assess the synchronization between spikes. Sec.~\ref{results} contains the results for the considered parameters; Sec.~\ref{conclusions} concludes the paper with some final considerations and remarks, and the Appendix reports some supplementary calculations cited in the main results section.

\begin{figure}[]
 \centering
 \includegraphics[width=0.99\linewidth]{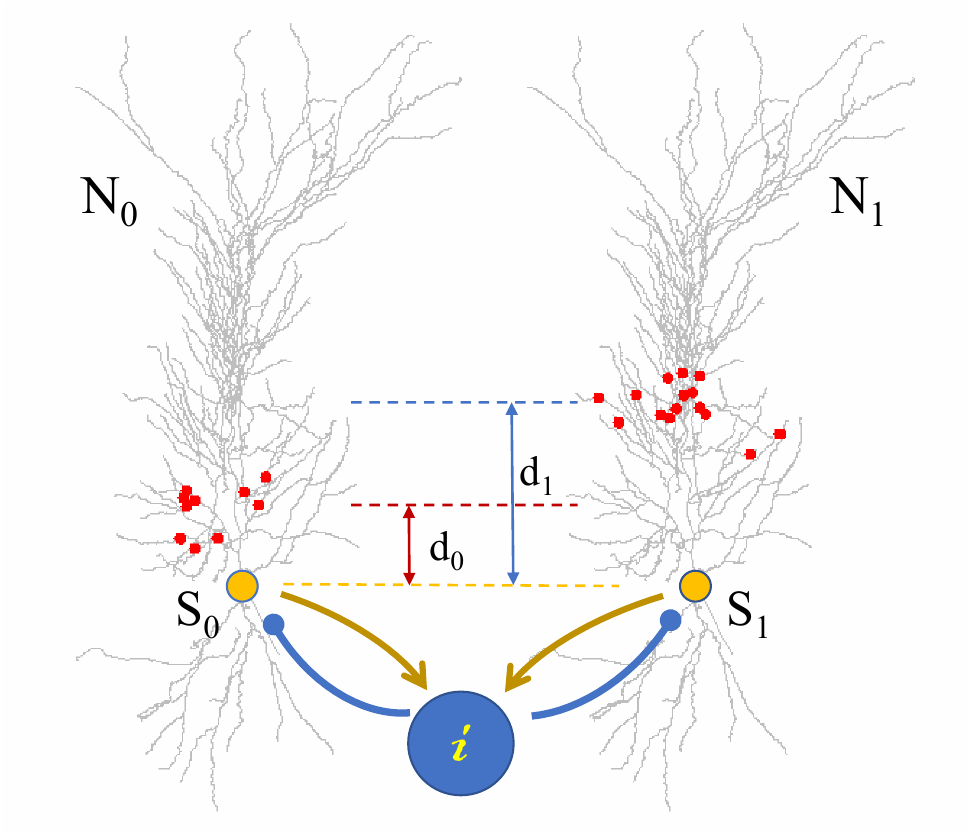}
 \caption{Scheme of the two copies of a full modeled CA1 neurons, specifically the neuron labeled mpg141017\_a1-2\_idC in Ref.~\cite{2018PLoS_migliore}, that receive inputs from the synapses (marked as red dots in the figure) randomly distributed at different distances from the soma along the apical dendrites. The neuron N$_1$, with a more distal distribution of synapses, receives the signals at a distance $d_1$ from the soma, while N$_0$ has the synapses in a more proximal region $d_0$.  Reciprocal inhibition was implemented with an inhibitory synapse placed on both S$_0$ and S$_1$, activated by the action potentials elicited in either neuron.}
\label{Neurons}
\end{figure}

\section{Methods} \label{methods}
%
There are many factors that can modulate the response of individual neurons to similar inputs. However, it should be noted that in spite of the different morphological and electrophysiological properties, there is experimental evidence showing in vivo that neurons may have common inputs that result in differentially synchronized responses during cognitive functions (see e.g. Ref.~\cite{2019Prescott}). In this work, we focus on one of the factors modulating this synchronization, i.e. synaptic inputs location, eliminating the influence of all the other factors by using identical, but otherwise detailed, neurons. For this purpose we used two identical neurons, each one receiving an ensemble of synaptic inputs clustered at different dendritic distance from the soma but synchronously activated. More specifically, the synapses on the first neuron are fixed at a proximal location, whereas synapses on the second neuron are placed at a variable (distal) location, as explained below. Figure~\ref{Neurons} shows the scheme of the chosen hippocampal neuron model, which consists of a morphologically and biophysically accurate structure with intrinsic electrophysiological properties consistent with experimental findings~\cite{2018PLoS_migliore}. The presynaptic signals consist of current pulses distributed in time according to a Poisson law. The source, which is identical for both neurons, is connected to each target by $N_{\rm syn}$ excitatory synapses. The two sets of synapses are located as follows: the proximal ones at a fixed average distance $d_0=100\,{\rm \mu m}$ from the soma on neuron $N_0$; the other set on $N_1$ at a variable average distance $d_1 \in [100,300]\,{\rm \mu m}$. The synapse distance is computed along the shortest path to the soma and is varied in each simulation according to a uniform random distribution in a window of $20\,\rm{\mu m}$ with $d_0$ and $d_1$ as ensemble mean with standard deviation $\sigma_d \approx 0.6\,\rm{\mu m}$.

We have chosen $N_{\rm syn}=20$, which represents a reasonable number of active synapses at any given time. This number is arbitrary and useful to provide a suitable intensity of stimulus producing spikes in the soma in our simulations. At the same time, it is sufficient to ensure a reasonable variety of random locations in the dendrites. The intensity of the signal can then be modulated by adjusting the excitatory synaptic weights.

To smooth the fluctuations in the outputs, the values were averaged over $N_{\rm exp}=100$ equivalent realizations by changing the initial random seed in each of them. Each simulation runs for a fixed time $t_{\rm sim} = 1\,$s.

\subsection{Equations and model details.}
The dynamics is based on the Hodgkin-Huxley (HH) set of equations, implemented in the NEURON package \cite{1997HinesCarnevale}, openly released by the Yale University. 

The realistic morphology requires the consideration of a large number of channels able to take into account the experimental features of CA1 pyramidal cells, against which the channel configuration and distribution have been thoroughly validated, as discussed in recent works on hippocampal neurons~\cite{2010Ascoli,2010Morse}. The complete set of active membrane properties includes the essential sodium (Na) current, four types of potassium (K$_{\rm DR}$, K$_{\rm A}$, K$_{\rm M}$, and K$_{\rm D}$), three types of calcium (CaN, CaL, CaT), the non-specific current I$_{\rm h}$, and two types of Ca-dependent K$^+$ currents, K$_{\rm Ca}$ and C$_{\rm agk}$. All dendritic compartments show a uniform distribution of channels, except for K$_{\rm A}$ and I$_{\rm h}$ which are known to increase linearly with distance from the soma in pyramidal cells~\cite{1999Hoffman,1999Magee}. The values for the peak conductance of each channel were optimized independently in each type of neuronal compartment (soma, axon, basal and apical dendrites), with a difference of one order of magnitude~\cite{2018PLoS_migliore} while the CA1 hippocampal neuron structure (see Ref.~\cite{2008Spruston} for a review) is optimized to ensure its realistic response to stimuli~\cite{2018PLoS_migliore}.

The excitatory post-synaptic currents are modeled as a double exponential with fixed rise time $\tau_1 = 3\, {\rm ms}$ and inactivation/decay time $\tau_2 \in [25, 50,75,100]\,{\rm ms}$ and the inhibitory current is modeled as a single exponential with decay time $\tau_{\rm I,d}=30\,{\rm ms}$. The reversal potentials are $E_{\rm r, E}=0$ and $E_{\rm r, I}=-80$, for the excitatory and the inhibitory cases, respectively.

The source of the excitatory synapses is a pulsed current $I$ generated by a Poissonian distribution of signals with mean input rates $f_{\rm inp} \in [25, 40, 60, 75]\,{\rm Hz}$.  All $N_{\rm syn}$ synapses of the two neurons receive exactly the same current, being connected to the same source $I$. In this way, the differences in the signal arriving at the somas are due only to the different localization of the synapses along the dendrites of the two neurons. We will see that even in the case of same mean distance from the somas ($d_1=d_0$), the correlation $\langle C_{\rm R} \rangle$, since it corresponds to different {\emph random} synapse localizations, is strongly reduced with respect to perfect synchronization. In fact, the ``perfect" case with correlation $\langle C_{\rm R}\rangle_{\rm ideal}=1$ is obtained only for {\emph identical} synapse distributions in the two neurons (even if they are changed in the different realizations), which are then trivially and expectantly maximally synchronized. Because of this significant difference, it becomes clear that synchronization is a very delicate phenomenon that deserves to be thoroughly understood in realistic neural systems.

The signals received at the synapses cause depolarization of the membrane potential that propagate in the dendrites following their individual paths to the soma and summing their contributions. Finally, they generate an action potential in the center of the soma, where the voltage is detected.

The action of the interneuron has been simulated in the form of reciprocal inhibitory synapses in both somas, creating a forward and backward inhibitory mechanism~\cite{2010Bazhenov}. The intensity of the inhibition is modulated by the weight of the synapses, which ranges from 0 to $0.06\,{\rm nS}$, according to a realistic inhibition weight for the synapses of an interneuron~\cite{2001Megias}.
In all simulations, we used an effective synaptic delay time of $\tau_{\rm sd}=2\,{\rm ms}$ to mark the onset of inhibition.

\subsection {Phase synchronization model.}
The correlation between the spikes in the two neurons is studied by means of a novel measure $c$, which modifies an equivalent measure already presented in~\cite{2011Perez}. The idea is to define the phase of a spiking neuron as~\cite{1997Pikovsky}:
\be
	\phi_{\rm i}(t) = 2\pi\frac{t-t_{\rm k}}{t_{\rm k+1,i}-t_{\rm k,i}},
\ee
where the index $i=0,1$ indicates the neuron $N_i$, and the time $t_{\rm k,i}$ is the time of the $k$-th spike of $N_i$. Thus, at time $t$, the neuron lays at a certain fraction of the total time between two successive spikes, the latter covering a phase between 0 and $2\pi$. This allows us to define the time average of the cosine of the phase difference between the two neurons in a single trajectory $j$ as 
\be
	c_{\rm j} = \frac{1}{t_{\rm fin}-{\rm t_{in}}} \int_{t_{\rm in}}^{t_{\rm fin}} \cos (\phi_1(t) - \phi_0(t)) \,dt,
\ee
where $t_{\rm fin}-t_{\rm in}$ defines the maximum window used for time averaging where the phase is well defined for both neurons.
This measure is normalized in the interval $[-1,1]$, where the value $-1$ represents the anticorrelated spiking (phase difference equal to $\pi$) and the full correlation is given by $+1$ (phase difference 0). The random distribution of the phases is then revealed by the correlation value $c_{\rm j} = 0$ (mean phase difference $\pi/2$).

Due to the stochastic nature of the spiking, this measure is then averaged over the number of simulated realizations $N_{\rm exp}$:
\be
	\langle C_{\rm R} \rangle = \frac{1}{N_{\rm exp}} \sum_{j=1}^{N_{\rm exp}} c_{\rm j},
	\label{cmmean}
\ee
This spiking synchronization measure, or others of the same kind, can be easily generalized to neuron synchronization in large networks~\cite{2011Perez}.

This approach allows us to consider the spike times as the only variable involved in the synchronization process. In contrast, the continuous cross-correlation between functions, often used in these contexts, would include the behavior of the entire trace, providing a more general measure rather than the more specific information on which we are interested, i.e. co-occurrence of spikes.

\section{Results} \label{results}
The spike activity at the soma of both neurons has been recorded for different input rates $f_{\rm inp}$, inactivation time of excitatory synapses $\tau_2$ and excitatory weight $sw$, as a function of the distance parameter $d_1$ and the inhibition weight $iw$.

\begin{figure}[t]
 \centering
 \includegraphics[width=\linewidth]{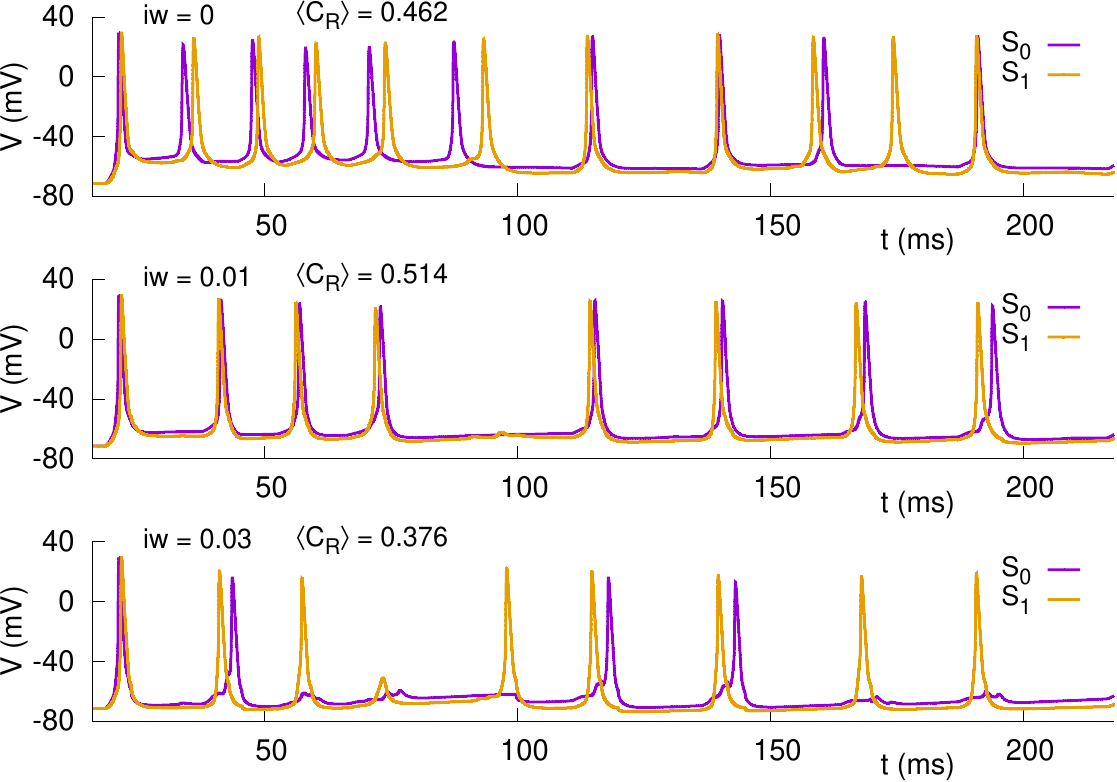}
 \caption{Time window of the evolution of the two neurons spiking for different inhibition weights, namely $iw=0, 0.01, 0.03\,{\rm nS}$. The trajectories labeled $S_0$, $S_1$ represent the depolarization potential in somas $S_0$ and $S_1$, respectively. The plot clearly shows the increase of synchronization at the optimal inhibition value $iw^*\approx0.01\,{\rm nS}$. The parameters are here: Synapse distances from the soma in $N_1$, $d_1=175\,{\rm \mu m}$, frequency input $f_{\rm inp}=60\,{\rm Hz}$, excitatory weight $sw=5\times 10^{-4}\,{\rm nS}$, inhibition delay $\tau_{\rm sd}=2\,{\rm ms}$ and inactivation time $\tau_2=50\,{\rm ms}$. The distance from the soma of the synapses in the 1$^{\rm st}$ neuron $N_0$ is kept fixed at the average value $d_0=100\,{\rm \mu m}$.}
\label{traj_ws5e-4_iw}
\end{figure}

\begin{figure}[t]
 \centering
 \includegraphics[width=\linewidth]{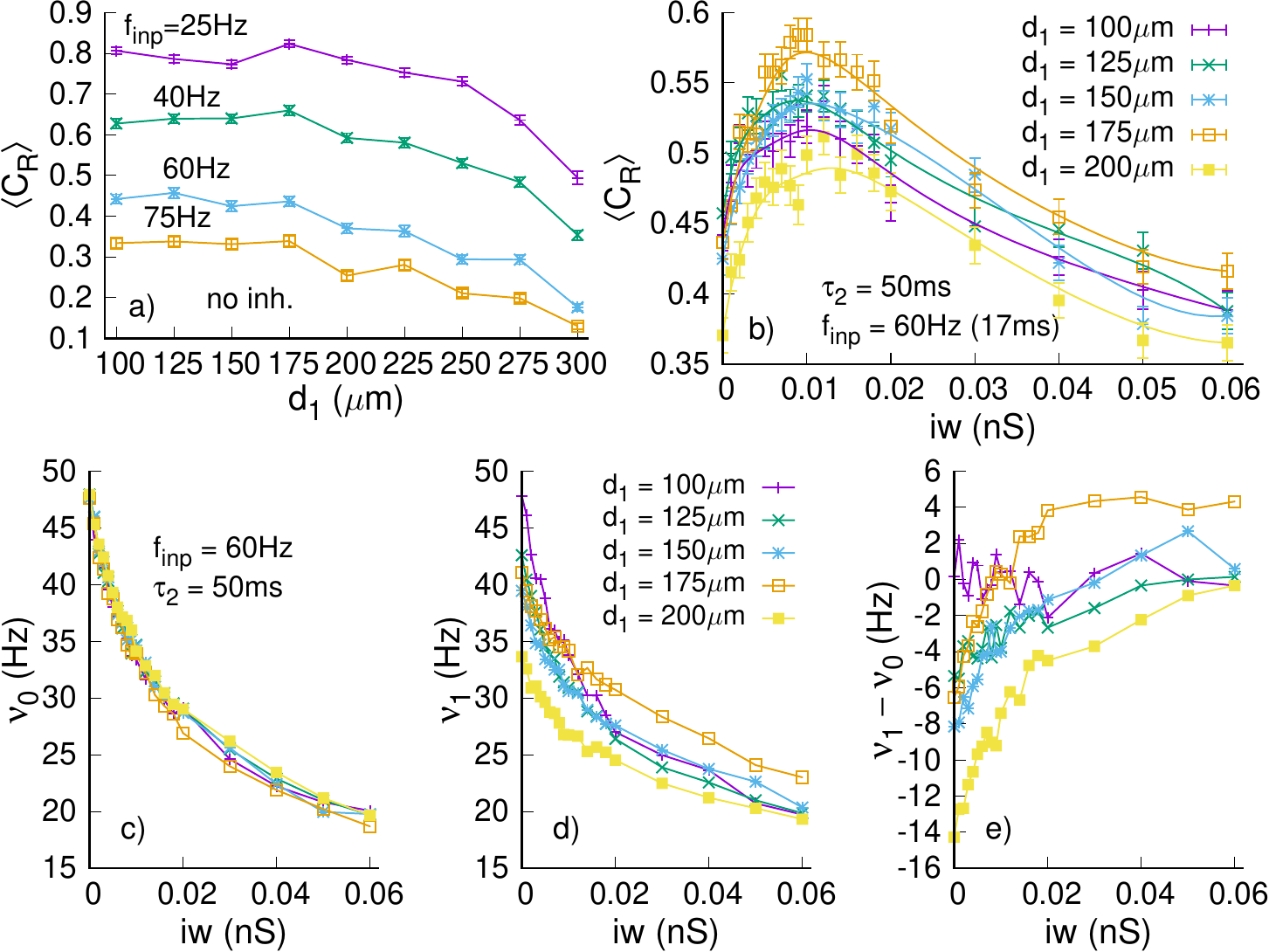}
 \caption{Panel a) Spiking correlation as a function of the distance $d_1$ with $iw=0$ for different input rates. Panel b) Spiking correlation as a function of the inhibition weight for different synapses distances from the soma of the second neuron ($d_1$). The inhibition induces a nonmonotonic behavior of $\langle C_{\rm R}\rangle$ for all the distances revealing an optimal inhibitory weight for the neurons to synchronize. Bottom panels: Mean firing response of the two neurons $N_0$ (panel c)) and $N_1$ (panel d)) and difference of the two spiking rates $\nu_1-\nu_0$ (panel e)). The other parameters are the same as Fig.~\ref{traj_ws5e-4_iw}.}
\label{Synch_iw_f}
\end{figure}

Figure~\ref{traj_ws5e-4_iw} shows some examples of spiking trajectories. The panels exhibit three cases with different inhibition weights $iw$ at frequency input $f_{\rm inp}=60\,{\rm Hz}$. The synchronization between spikes $\langle C_{\rm R}\rangle$ is maximal at the value $iw^* = 0.01\,{\rm nS}$, as we will show in Fig.~\ref{Synch_iw_f}, by averaging over $N_{\rm exp}$ simulations. 

A counterintuitive phenomenon revealed by the simulations is the possibility that distal synapses may elicit a somatic spike earlier than proximal ones. Fig.~\ref{traj_ws5e-4_iw} shows several events where this phenomenon occurs: The spiking of $S_1$, whose synapses are distributed at $d_1=175,{\rm \mu m}$ from the soma, sometimes precedes the spiking of $S_0$, whose synapses are closer to the soma ($d_0=100\,{\rm \mu m}$). This apparently random behavior persists even when the inter-spike interval (ISI) is smaller (higher frequency response of the neuron) or when other parameters are changed.
  At first glance, these findings seem to contradict the results reported in Ref.~\cite{2010Schultheiss} where the current injected into distal synapses produces a smaller spiking anticipation then the current injected into closer synapses, in the context of a full morphological model. A similar effect is reported in Ref.~\cite{1998Crook} within a schematized model, with the difference that the phase response can be modulated by the presence of active channel current along the dendrites.
 However, unlike our work, which aims to emulate realistic neuronal behavior, these studies examine the phase response under the application of a non-realistic constant current that causes neurons to spike regularly and continuously. Therefore, a direct comparison with those results cannot be made.
 
 One possible explanation for this erratic behavior is the specificity of dendritic signal propagation. Distal sites are narrower than proximal ones, and it has been reported that the local voltage response to the same current is higher in narrow dendrites than in wider ones~\cite{2010Schultheiss}. In addition, voltage propagation may be more efficient again in narrow dendrites than in wider ones~\cite{2005MiglioreAscoli}. So, we try to explain this phenomenon in the following way: The depolarization propagating towards the dendrite edges, if it comes from narrow sections, that is, from those synapses located very close to the dendrite arbors, quickly rebounds at the dendrite terminals. The rebound depolarization may have the possibility to sum with the direct signal toward the soma, generating an increased potential depolarization and contributing to the anticipation of spikes in the soma. This mechanism follows a pattern similar to that described by Schultheiss et al. in Ref.~\cite{2010Schultheiss}.

Certainly, this hypothesis would require a more extensive and dedicated study for a proper validation. However, this is beyond the scope of this manuscript.

The spiking trajectories plotted in Fig.~\ref{traj_ws5e-4_iw} show that for almost all inhibitory weights used, occasionally only one of the neurons spikes. This unpredictable event (see panel $iw=0\,{\rm nS}$) can be amplified by the action of the interneurons (panel $iw=0.03\,{\rm nS}$). Obviously, this kind of events contributes to the reduction of the synchronization properties, which leads to the well visible nonmonotonic behavior in $\langle C_{\rm R} \rangle$ as a function of the inhibition weight $iw$ (see Fig.~\ref{Synch_iw_f}b)).

The above observations in a single depolarization trajectory, are affected by the intrinsic stochasticity of the synapse distribution which is characteristic of this study. Then, the ensemble average of different trajectories can better elucidate the average neural behavior.
Figure~\ref{Synch_iw_f}a) shows the mean spiking correlation $\langle C_{\rm R} \rangle$ for uninhibited dynamics ($iw=0\,{\rm nS}$) as a function of distance $d_1$ for different values of $f_{\rm inp}$. As already mentioned, even at the same mean distance ($d_1=d_0 = 100\,{\rm \mu m}$) the correlation decreases significantly (see the values at $d_1=100$ in panel a)) compared to the ideal/trivial case ($\langle C_{\rm R} \rangle_{\rm ideal}=1$). This decrease is due to the stochastic location of the synapses and the complex structure and functionality of the apical dendritic branches of the realistic neuron.
The effect on the correlations remain approximately constant up to $d_1\approx 175\,{\rm \mu m}$, and then decreases rapidly with increasing distance.

The distance value $d_1^*\approx 175\,{\rm \mu m}$ results in a higher $\langle C_{\rm R} \rangle$ even across different values of the inhibition weight $iw$, as shown in panel b) at the input frequency $f_{\rm inp}=50\,{\rm Hz}$, and excitatory decay $\tau_2=50\,{\rm ms}$. The curves exhibit a pronounced nonmonotonic behavior as a function of $iw$ for all the distances, clearly identifying the maximum inhibition value at $iw^* \approx 0.01\,{\rm nS}$. This indicates the existence of an optimal inhibition weight at which spiking synchronization peaks.
This result may not be obvious. In fact, the well-established constructive mechanism of synchronization induced by the inhibitory action of interneurons in the postsynaptic potential (IPSP)~\cite{Jasper,2010Bazhenov,1996Buzsaki}, is valid for periodic signals. This case can be reproduced in the laboratory by applying a continuous current to a single neuron, which induces periodic neuronal firing. In this direction, we have carried out a series of simulations under {\it current clamp} conditions, plus a pulsed synaptic stimulus at fixed frequency, for point-like somas, which essentially agree with the results reported in Ref.~\cite{1994Vreeswijk}. We verified the increase in synchronization by applying a suitable inhibitory signal in three cases: initially phase-shifted signals with the same constant current, different initial frequencies, phase-shifted signals due to delayed stimulation on a set of synapses. The effect of inhibition was able to induce synchronization in phase and frequency in all cases. All these results are presented in the figures~\ref{Dfini}, \ref{DI}, \ref{Dsyn} of the subsection \ref{subB} of the appendix. On the other hand, when only the Poisson pulse current was used as input to our neurons, the synchronization showed a non-trivial behavior. In particular, for short inactivation delays (i.e. $\tau_2 < 25\,{\rm ms}$), the correlation showed a rapid monotonic decrease with increasing inhibition weight. For longer inactivation delay times ($\tau_2 \ge 25\,{\rm ms}$) the correlation exhibited a similar nonmonotonic behavior as shown in Fig.~\ref{Synch_iw_f}b).

The increase in synchronization with inhibition then seems to be associated with a sufficiently long inactivation time $\tau_2$ ---typical of NMDA-mediated synapses~\cite{1997Haos,1999Kleppe,2004Caballero}--- which allows the inhibition to act during subsequent spikes, thus approaching the mechanism of continuous current supply. Under these conditions, interneurons in the hippocampal network reaffirm their importance in increasing the synchronization efficiency for some optimal value of their intensity. This behavior seems to be consistent with the reported action of fast AMPA-type glutamatergic synapses, characterized by short $\tau_2$ decay times, which usually tend to desynchronize rather than synchronize repetitive spike rings of coupled neurons~\cite{1994Vreeswijk}.

The bottom panels of Fig.~\ref{Synch_iw_f} show the firing rate response ($\nu_0$ and $\nu_1$) of the two neurons for different distances $d_1$ as a function of $iw$. We observe that inhibition induces a monotonic decrease in the spiking response of both neurons. However, while this trend is consistent over $d_1$ for neuron $N_0$ (see panel d)), differences emerge for neuron $N_1$ (panel c)). As a consequence, the firing rate difference $\nu_1-\nu_0$ shows interesting features worth highlighting. Panel e) shows a decreasing difference of the spiking activity between the two neurons with the inhibition weight, and even an increase of $\nu_1$ with respect to $\nu_0$ for some of the distances $d_1$. In fact, besides the clear monotonic behavior in both neurons with inhibition, $\nu_1$ shows a nonmonotonic trend with the distance $d_1$. Specifically, for $d_1=175\,{\rm \mu m}$, i.e. the value that gives the maximum correlation $\langle C_{\rm R} \rangle$, $S_1$ spikes more than $S_0$ for sufficiently high inhibition weights. Under these conditions, the spiking rate is higher in the neuron with distal synapses $N_1$ compared to $N_0$ with a clear crossing of the zero value, indicating equal spiking rate at this $iw$ value. As already mentioned, this behavior can be attributed to the highly structured dendritic distribution of the realistic neuron, which presents optimal distances for the depolarization to occur, due to optimal reflection mechanisms of the current traveling along the dendrites \cite{2005MiglioreAscoli}. The mechanism of synchronization at the crossing value resembles the mechanism of synchronization over long times in neurons with different firing rates that tend to eventually coincide because of their diverse decreasing trends~\cite{1995Hopfield,2000HopfieldBrody}.

\begin{figure}[tb]
\includegraphics[width=\linewidth]{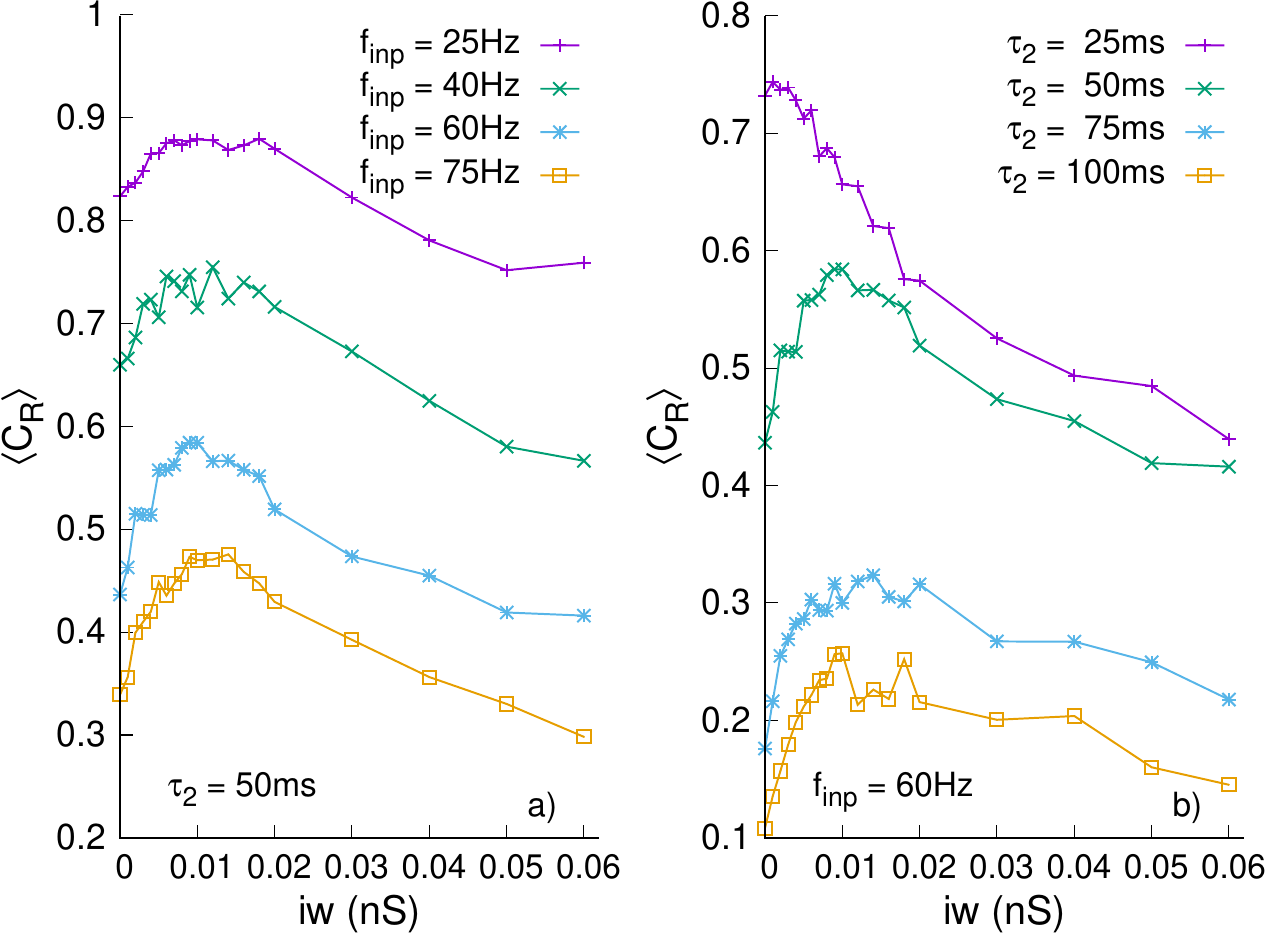}
\caption{Spiking correlation as a function of the inhibitory weight $iw$ for different input rates (panel a)) and different inactivation times (panel b)) for $d_1=175\,\mu {\rm m}$. In panel a) the presence of inhibition confirms the nonmonotonic behavior for all firing rates at the optimum inhibitory weight $iw^* \approx 0.01\,{\rm nS}$.  Panel b) shows how the curve for $\tau_2=25\,{\rm ms}$ misses the correlation increase with the inhibition, although the correlation is higher than at longer $\tau_2$.}
\label{Span_Tf}
\end{figure}
The optimal correlation corresponding to the distance value $d_1^*= 175\,{\rm \mu m}$ is also confirmed for all the other parameter ranges explored in this paper. This includes different input rates $f_{\rm inp}$ and different values of synapse inactivation time $\tau_2$. However, this value depends on the morphology used. In fact, some calculations with a different morphology (see Fig.~\ref{M2} of the appendix, subsection~\ref{subA}) show a nonmonotonic behavior of the correlation with inhibition, at an optimal inhibition value in the same range as the one shown in Figs.~\ref{Synch_iw_f} and \ref{Span_Tf}, but with a different distance $d_1^*$. This means that different neurons respond differently to the synapse location, due to their specific morphology. In all cases, we also observe a nonmonotonic behavior of the correlation with the distance $d_1$.

Figure~\ref{Span_Tf} shows, for $d_1= 175\,{\rm \mu m}$, the spiking correlation as a function of the inhibition weight $iw$, for different mean input rates (panel a)) and different inactivation times $\tau_2$ (panel b)).
In panel a) we observe that the lower the input rate, the higher the spiking correlation, while the optimal inhibition value $iw^*$ remains constant and the peak around the local maximum widens slightly. This behavior is attributed to the increase of spiking failures in the trajectories of one of the two somas with increasing the firing rate response, consistent with the increased time spread between the refractory period and the mean input period of the current pulses.

In panel b) we observe that for a fixed value of the input rate $f_{\rm inp}$, the correlation decreases as $\tau_2$ increases. As mentioned above, we can see that for $\tau_2 = 25\,{\rm ms}$ the maximum of the curve has disappeared resulting only in a monotonic decrease of the correlation. At high values of $\tau_2$, the correlation decreases because the spikes tend to overlap more during the refractory period, adding a randomizing contribution to the depolarization activity, and leading to an increasing number of failures. 

\begin{figure}[t]
 \centering
\includegraphics[width=0.9\linewidth]{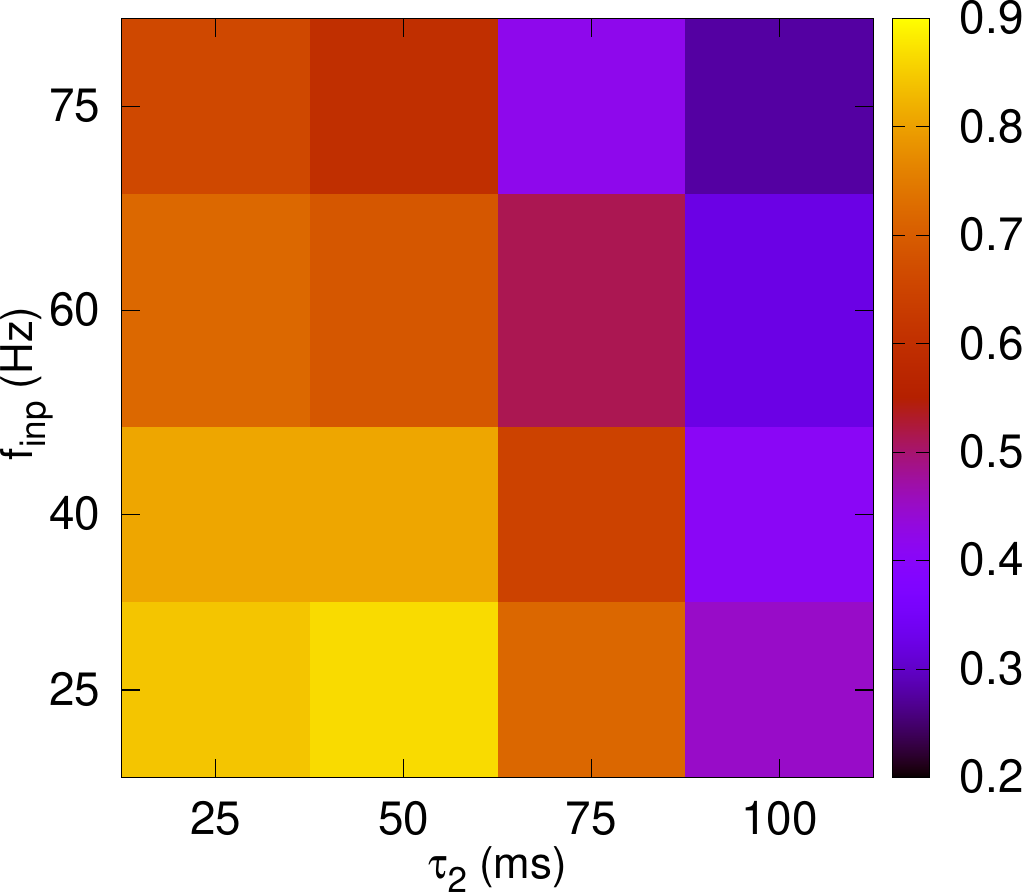}
 \caption{Spiking correlation  $\langle C_{\rm R} \rangle$ as a function of the input rate $f_{\rm inp}$ and the inactivation time $\tau_2$ for the distance $d_1 = 175\,{\rm \mu m}$ and inhibition weight $iw=0.01\,{\rm nS}$.}
\label{f-tau-map}
\end{figure}

The relationship between synaptic decay time and synchronization has also been discussed previously in a single compartment model in Ref.~\cite{1996Buzsaki}, where a constant current was injected to the system. Among other conditions, the authors identified a large enough ratio between the synaptic decay time ($\tau_2$) and the natural oscillation period $T$ of the neuron under the constant current for synchronization to occur.

To briefly summarize the above results, Fig.~\ref{f-tau-map} shows a color map depicting the spiking correlation as a function of input rate $f_{\rm inp}$ and inactivation time $\tau_2$, under the condition of maximum distance $d_1 = 175\,{\rm \mu m}$ and inhibition weight $iw=0.01\,{\rm nS}$. We observe that for certain input rates the highest $\langle C_{\rm R} \rangle$ value occurs at $\tau_2=50\,{\rm ms}$. However, for these cases, the local maximum of the curve with respect to $iw$ disappears, resulting in a monotonic behavior of the correlation with inhibition over all input rates considered.
\begin{figure}[b]
 \centering
\includegraphics[width=\linewidth]{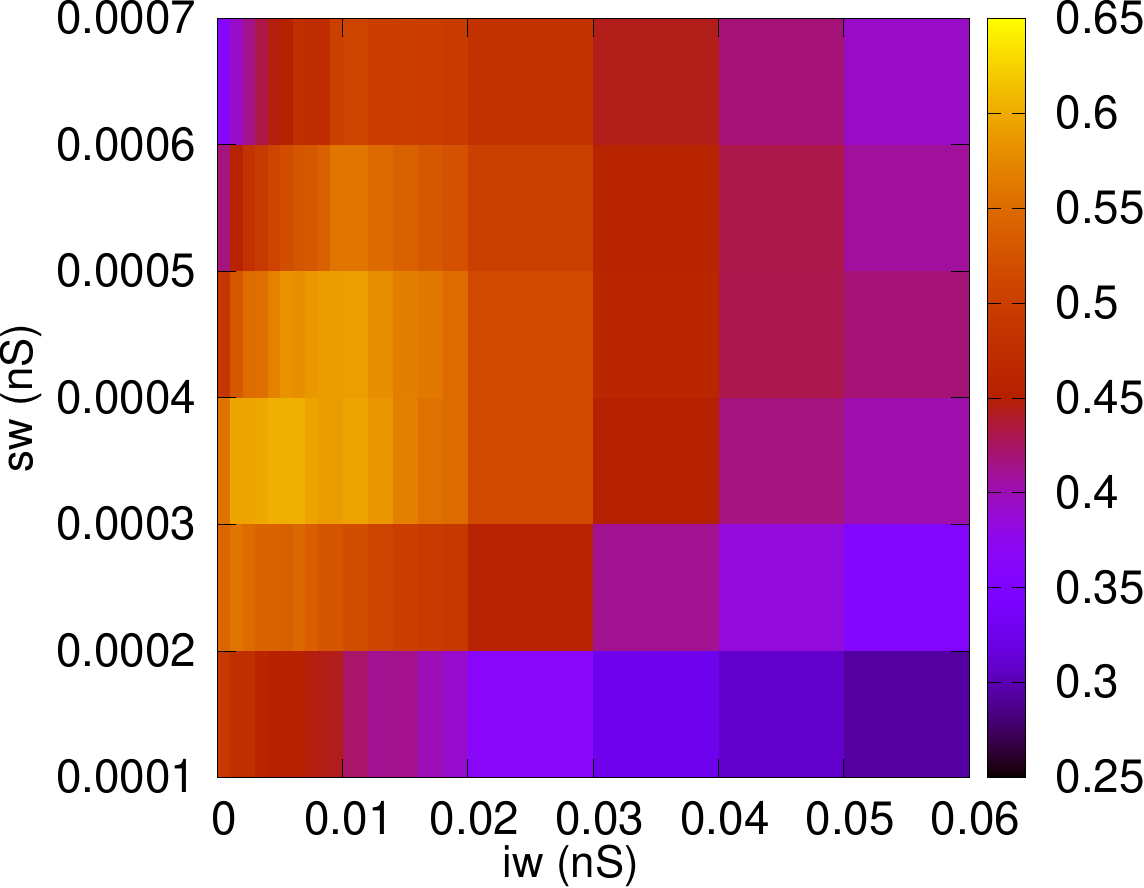}
 \caption{Spiking correlation  $\langle C_{\rm R} \rangle$ as a function of excitatory synapse weight $sw$ and inhibitory interneuron weight $iw$. The mean input rate is $f_{\rm inp}=60\,{\rm Hz}$, the inactivation time $\tau_2=50\,{\rm ms}$ and $d_1=175\,{\rm \mu m}$.}
\label{sw-iw-map}
\end{figure}

Finally, Fig.~\ref{sw-iw-map} shows a color map of the spiking correlation $\langle C_{\rm R} \rangle$  as a function of the excitatory weight $sw$ and the inhibition weight $iw$ for input rate $f_{\rm inp}=60\,{\rm Hz}$ and inactivation time $\tau_2=50\,{\rm ms}$. Once again we observe an optimal range of inhibition/excitation magnitudes that gives the highest correlation for the parameters considered. This confirms the delicate condition under which synchronization can occur, also in terms of excitatory weights.

\section{Summary and Conclusions.} \label{conclusions}
In this work, we study the synchronization behavior of two realistic CA1 neuron models under the input of Poissonian distributed current pulses with different mean input rates in the gamma range $f_{\rm inp}=25, 40, 60, 75 \,{\rm Hz}$. A novel spiking synchronization measure has been studied as a function of the inhibition weight of the interneuron connecting the two somas for different inactivation times of the excitatory synapses. The signals are assumed to originate from the CA3 neurons of the hippocampal trisynaptic path and arrive at different sites of the apical dendrites, thus modeling realistic dynamics inside the neurons.

No explicit delay has been imposed at the excitatory synapses. Its addition can mimic the time differences due to different paths arriving at the two neurons, thereby increasing the stochasticity of the Poissonian current signals and leading to a decrease in the spiking correlation $\langle C_{\rm R} \rangle$.

In our calculations we noticed the following:
\begin{itemize}

\item The spiking activity of the second neuron (with more distal synapses) exhibits a non-monotonic dependence on synapse localization.
\item The difference between the mean firing rate response of the two somas changes sign for certain distances at high inhibition weights, consistent with the aforementioned nonmonotonic behavior of spiking rate with distance.
\item The difference between the mean spiking rates tends to saturate at high inhibition weights.
\item The inhibitory activity leads to maximum synchronization at certain weights, provided that the inactivation time of excitatory synapses is sufficiently long, a condition that is met in NMDA-mediated synapses. In this sense, the model suggests that AMPA-mediated synapses, characterized by a short inactivation time, do not exhibit an increase in synchronization induced by inhibitory activity of interneurons in the case of Poisson distributed excitations.
\end{itemize}

The results of this study show that synchronization emerges from a nuanced interplay of favorable conditions, including specific neuron morphology, the presence of NMDA-mediated synapses, and dynamic parameters such as appropriate synaptic and inhibitory weights. Furthermore, we have shown here that low input rates enhance the spiking synchronization between neurons.

\section{Acknowledgments}
The authors acknowledge the grant PID2020-113582GB-I00 funded by MCIN/AEI/ 10.13039/501100011033, the support of the Aragon Government to the Recognized group `E36\_23R F\'isica Estad\'istica y no-lineal (FENOL)' and the grant from the Swiss National Supercomputing Centre (CSCS) under project ID ich011 and ich002 (to MM), and from the CINECA Consortium (Italy).
We also acknowledge a contribution from the Italian National Recovery and Resilience Plan (NRRP), M4C2, funded by the European Union – NextGenerationEU (Project IR0000011, CUP B51E22000150006, EBRAINS-Italy), and the funds of the European Union-NextGenerationEU through the spanish Ministerio de Universidades (BOA 139 (31185) 01/07/2021).

\section{Appendix.} \label{Appendix}

\subsection{Alternative morphology}  \label{subA}
\begin{figure}[h]
 \centering
\includegraphics[width=\linewidth]{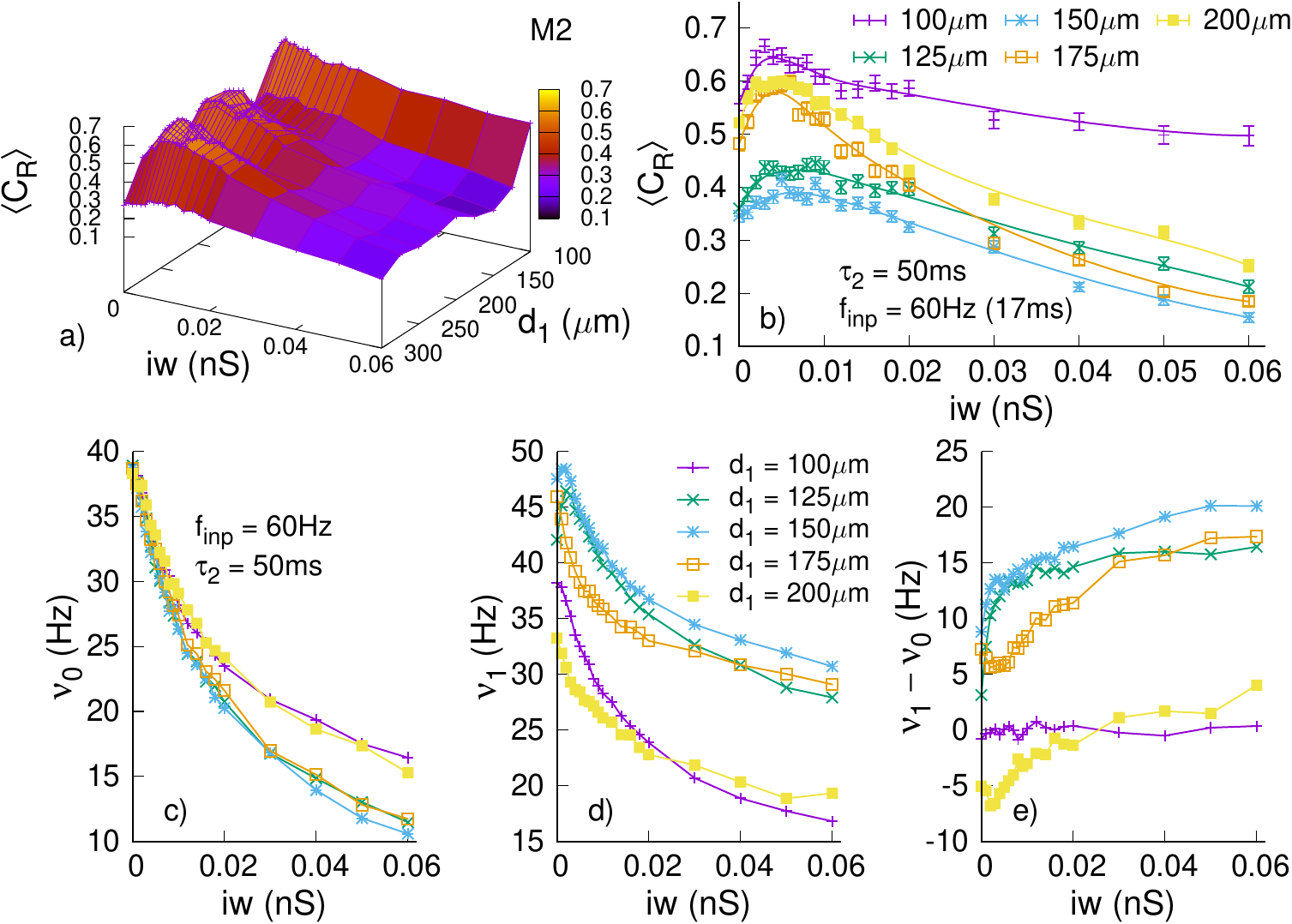}
 \caption{Spiking correlation  $\langle C_{\rm R} \rangle$ as a function of inhibition weight $iw$ and the synapse distance $d_1$ at the input rate $f_{\rm inp}=60\,{\rm Hz}$ and $\tau_2=50\,{\rm ms}$, by using a different morphology (mpg141208\_B\_idA~\cite{2018PLoS_migliore}). We can see in panel b) that the maximum correlation is obtained at an optimal inhibition weight, whose higher value is obtained for a closer synapse distribution $d_1^*=100\,{\rm \mu m}$ then that shown in Fig.~\ref{Synch_iw_f} ($d_1^*=175\,{\rm \mu m}$).}
\label{M2}
\end{figure}

To check the robustness of the results, we use here another morphology, specifically the one labeled mpg141208\_B\_idA in Ref.~\cite{2018PLoS_migliore}. Fig.~\ref{M2} shows the equivalent of Fig.~\ref{Synch_iw_f} in panels b) to e), which shows that the qualitative behavior is preserved: we can see the clear nonmonotonic behavior of $\langle C_{\rm R} \rangle$, but a different optimal inhibition value than before: $iw \approx 0.005\,{\rm nS}$. Again, the behavior of the spiking correlation as a function of the distance of the synapses remains nonmonotonic. In this case, the maximum correlation is obtained for proximal synapses $d_{\rm 1, M2}^*=100\,{\rm \mu m}$ instead of the previous value $d_1^*=175\,{\rm \mu m}$, closely followed by those for distal synapses $d_1=175\,{\rm \mu m}$ and $d_1=200\,{\rm \mu m}$. It becomes clear that the quantitative behavior of the synchronization features is not trivially predictable on the basis of synaptic distances from the soma alone, when using different morphologies. 

\subsection{Synchronization by inhibition} \label{subB}
We have performed a series of examples of the effects of inhibition on the synchronization of two point-like neurons. The simple model used here is based on passive membrane properties and the presence of K$^+$ and Na$^+$ channels only. Both neurons are charged with a constant current $I_{\rm C}=1$pA, i.e. the simulations are run in {\it current clamp} mode, so that a constant spiking frequency is induced on the two neurons. A second input is given by the synaptic current $I_{\rm syn}$, which stimulates the synapses on both somas with periodic pulses of frequency $f_{inp}=50\,{\rm Hz}$. 
The post synaptic potentials are modeled with a double exponential function for the excitatory synapses with rise time $\tau_1=0.5\,{\rm ms}$ and decay time $\tau_2=10\,{\rm ms}$, while the inhibitory synapse is modeled as a single exponential with decay time $\tau_{\rm I2}=30\,{\rm ms}$. The inhibition is activated by the soma's spikes and always acts on both of them at the same time.

\begin{figure}[t]
 \centering
 \vspace{3mm}
\includegraphics[width=\linewidth]{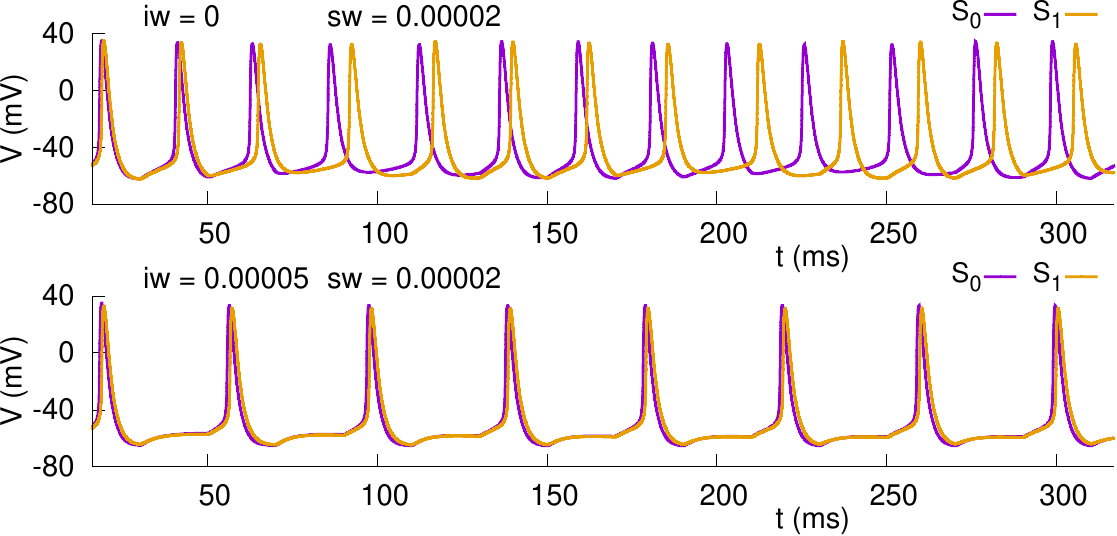}
 \caption{Membrane potential of the two neurons ($S_0$ and $S_1$). The upper panel shows the uninhibited dynamics $iw=0\,{\rm nS}$ with different spiking frequencies in the two somas due to the application of the a current attenuated  by a factor 0.95 in $S_1$ with respect to $S_0$. The bottom panel shows the synchronization induced by the inhibition weight $iw=5\times 10^{-5}\,{\rm nS}$. The excitatory synaptic weight is $sw=2\times10^{-5}\,{\rm nS}$.}
\label{DI}
\end{figure}

\begin{figure}[b]
 \centering
\includegraphics[width=\linewidth]{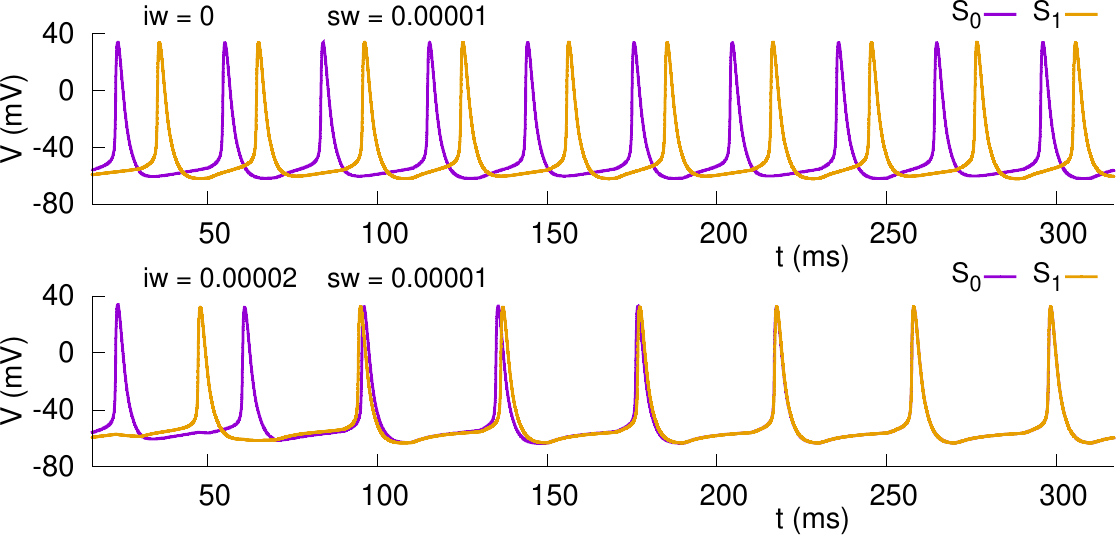}
 \caption{Membrane potential of the two neurons ($S_0$ and $S_1$). The upper panel shows the uninhibited dynamics with dephased signal response due to a delayed excitatory current pulse $I_{\rm syn}$ at the synapses of 20ms in $S_1$ with respect to $S_0$. The bottom panel shows the synchronization induced by the inhibition weight $iw=2\times 10^{-5}\,{\rm nS}$. The excitatory synaptic weight is $sw=10^{-5}\,{\rm nS}$.}
\label{Dsyn}
\end{figure}

We show here three examples of different kinetic conditions of the neurons, and the synchronizing effect of inhibition on them.
\begin{itemize}
\item
 The first example concerns depolarization response of the somas under an activating current on neuron $S_1$ with an initial delay of 30ms with respect to $S_0$. The phase difference in the response is maintained for a long time for uninhibited somas. (See top panel of Fig.~\ref{Dfini})
\begin{figure}[h]
 \centering
\includegraphics[width=\linewidth]{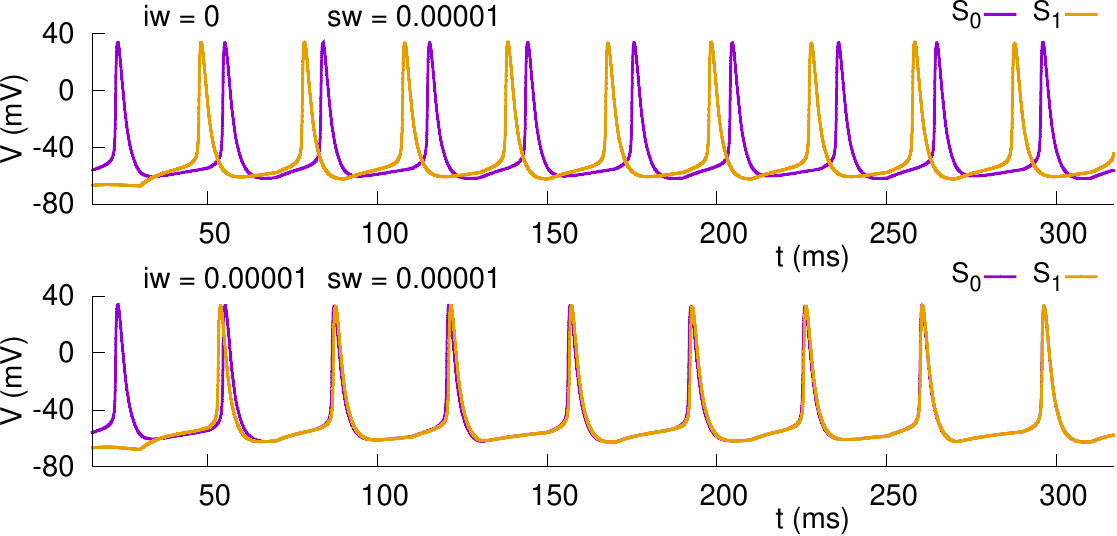}
 \caption{Membrane potential of the two neurons ($S_0$ and $S_1$). The top panel shows the uninhibited dynamics with dephased signals due to the delayed application of the constant current $I_{\rm C}$ in $S_1$. The bottom panel shows the synchronization induced by the inhibition weight $iw=10^{-5}\,{\rm nS}$. The excitatory synaptic weight is $sw=10^{-5}\,{\rm nS}$.}
\label{Dfini}
\end{figure}

\item
 In the 2$^{nd}$ example, we observe a different frequency response of the neurons, due to the application of an attenuated constant current in the soma $S_1$ by a factor 0.95 with respect to $S_0$. (See top panel of Fig.~\ref{DI}).
\item
 The 3$^{rd}$ example represents the phase difference in the neuron's response due to the application of the pulsed current  $I_{\rm syn}$ in soma $S_1$ delayed by a constant time of 20ms with respect to $S_0$.  (See Fig.~\ref{Dsyn}).

\end{itemize}

In all three cases, the application of an appropriate level of inhibition has the effect of reducing the frequency of the spike responses and synchronizing the two neurons in great detail. The effect is clearly visible in the lower panel of each of the three figures \ref{Dfini}, \ref{DI} and \ref{Dsyn}.



\begin{thebibliography}{99} 

\bibitem{1993Okeefe} O'Keefe J \& Recce ML.
 Phase relationship between hippocampal place units and the EEG theta rhythm.
  {\em Hippocampus} {\bf 3} 317 (1993)

\bibitem{2015Fries} Fries P.
   Rhythms for Cognition: Communication through Coherence.
  {\em Neuron} {\bf 88} 220 (2015)

\bibitem{1995Mainen} Mainen ZF \& Sejnowski TJ.
   Reliability of spike timing in neocortical neurons.
  {\em Science} {\bf 268} 1503 (1995)

\bibitem{1995Lisnan} Lisman JE \& Idiart MA.
   Storage of 7 +/- 2 short-term memories in oscillatory subcycles.
  {\em Science} {\bf 267} 1512 (1995)

\bibitem{1995Hopfield} Hopfield JJ.
   Pattern recognition computation using action potential timing for stimulus representation.
  {\em Nature} {\bf  376} 33 (1995)
  
\bibitem{2000HopfieldBrody} Hopfield JJ \& Brody CD. 
   What is a moment? “Cortical” sensory integration over a brief interval.
 {\em Proc. Natl. Acad. Sci. USA} {\bf 97} 13919 (2000)



\bibitem{2003Izhikewich} Izhikewich EM.
 Simple Model of Spiking Neurons.
  {\em IEEE Transactions on Neural Networks} {\bf 14} 1569 (2003).

\bibitem{1975Kuramoto} Kuramoto Y,
   Self-entrainment of a population of coupled nonlinear oscillators.
  {\em Lect. Notes Phys.} {\bf 39} 420 (1975).

\bibitem{2005Ritort} Acebr\'on JA, Bonilla LL, P\'erez-Vicente CJ, Ritort F, and Spigler R.
 The kuramoto model: A simple paradigm for synchronization phenomena.
  {\em Rev. Mod. Phys.} {\bf 77} 137 (2005).

\bibitem{1999Abbott-LIF}  Abbott LF.
  Lapicque's introduction of the integrate-and-fire model neuron (1907).
    {\em Brain Research Bulletin}. {\bf 50}(5–6), 303 (1999). doi:10.1016/S0361-9230(99)00161-6. 

\bibitem{2005Brette} Brette R, Gerstner W.
 Adaptive Exponential Integrate-and-Fire Model as an Effective Description of Neuronal Activity. 
 {\em J. Neurophysiol.} {\bf 94}(5) 3637 (2005). doi:10.1152/jn.00686.2005.

\bibitem{2021Gorski} G\'orski T, Depannemaecker D, Destexhe A.
 Conductance-Based Adaptive Exponential Integrate-and-Fire Model. 
 {\em Neural Comput.} {\bf 33}(1) 41 (2021). doi:10.1162/neco\_a\_01342.

\bibitem{2023Marasco}  Marasco A, Spera E, De Falco V et al.
 An Adaptive Generalized Leaky Integrate-and-Fire Model for Hippocampal CA1 Pyramidal Neurons and Interneurons.
     {\em Bull Math Biol}. {\bf 85} 109 (2023). https://doi.org/10.1007/s11538-023-01206-8.

\bibitem{1952HH}  Hodgkin AL, Huxley AF. 
   A quantitative description of membrane current and its application to conduction and excitation in nerve.
     {\em J. Physiol.}. {\bf 11} 500 (1952).

\bibitem{2015LiGirault} Li L, Gervasi N and Girault JA. 
  Dendritic geometry shapes neuronal cAMP signalling to the nucleus
  {\em Nature Communications} {\bf 6} 6319 (2015).

\bibitem{2002Migliore} Migliore M, Sheperd GM. 
 Emerging rules for the distributions of active dendritic conductances. 
  {\em Nature Reviews Neuroscience} {\bf 3} 362 (2002).

\bibitem{1997HinesCarnevale} Hines ML, Carnevale NT.  
 The NEURON simulation environment. 
  {\em Neural Comput.} {\bf 9} 1179 (1997).

\bibitem{2008Spruston} Spruston N.
 Pyramidal neurons: dendritic structure and synaptic integration.
  {\em Nature Reviews Neuroscience} {\bf 9}, 206-221 (2008).

\bibitem{2018PLoS_migliore} Migliore R, Lupascu CA, Bologna LL, Romani A, Courcol J-D, Antonel S, et al.
 The physiological variability of channel density in hippocampal CA1 pyramidal cells and interneurons explored using a unified data-driven modeling workflow. 
  {\em PLoS Comput Biol} {\bf 14}(9): e1006423. (2018) https://doi.org/10.1371/journal.pcbi.1006423.
  
\bibitem{2005MiglioreAscoli} Migliore M, Ferrante M, Ascoli GA. %
 Signal propagation in oblique dendrites of CA1 pyramidal cells.
  {\em J. Neurophysiol.} {\bf 94} 4145 (2005). 

\bibitem{2001Megias} Meg\'ias M, Emri Zs, Freund TF and Guly\'as AI. %
 Total number and distribution of inhibitory and excitatory synapses on hippocampal ca1 pyramidal cells. 
  {\em Neuroscience} {\bf 102}(3) 527 (2001). 

\bibitem{2011Perez} P\'erez T, Garcia GC, Egu\'iluz VM, Vicente R, Pipa G, Mirasso C. 
 Effect of the Topology and Delayed Interactions in Neuronal Networks Synchronization.
  {\em  PLoS ONE} {\bf 6} e19900 (2011).

\bibitem{1997Pikovsky} Pikovsky A, Rosenblum M, Osipov G, Kurths J.
 Phase synchronization of chaotic oscillators by external driving.
  {\em  Physica D} {\bf 104} 219 (1997).

\bibitem{2010Bazhenov} Bazhenov M, Stopfer M.
 Forward and back: Motifs of inhibition in olfactory processing.
  {\em Neuron} {\bf 167} 357 (2010).
 
\bibitem{Jasper} Noebels JL, Avoli M, Rogawski MA,Olsen RW, Delgado-Escueta AV.
 {\em Jasper's Basic Mechanisms of the Epilepsies. Fourth edition}.
  Oxford University Press, USA (2012), pagg. 170-171.

\bibitem{1997Haos} H\'aos N, Mody I.  
   Synaptic Communication among Hippocampal Interneurons: Properties of Spontaneous IPSCs in Morphologically Identified Cells.
   {\em Journal of Neuroscience} {\bf 17}(21), 8427 (1997).

\bibitem{1999Kleppe} Kleppe IC, Robinson HPC.
   Determining the Activation Time Course of Synaptic AMPA Receptors from Openings of Colocalized NMDA Receptors.
   {\em Biophys. J} {\bf 77}, 1418 (1999).

\bibitem{2004Caballero} Vargas-Caballero M and Robinson HPC.   
   Fast and Slow Voltage-Dependent Dynamics of Magnesium Block in the NMDA Receptor: The Asymmetric Trapping Block Model.
   {\em Journal of Neuroscience} {\bf 24}(27), 6171 (2004).

\bibitem{2005Stuart} Gulledge AT, Kampa BM,  Stuart GJ. 
  Synaptic Integration in Dendritic Trees
  {\em J. Neurophysiol.} {\bf 64} 75 (2005).
  
\bibitem{2010Schultheiss} Schultheiss NW, Edgerton JR, and Jaeger D.   
   Phase Response Curve Analysis of a Full Morphological Globus Pallidus Neuron Model Reveals Distinct Perisomatic and Dendritic Modes of Synaptic Integration.
   {\em The Journal of Neuroscience} {\bf 30}(7), 2767 (2010).

\bibitem{2010Ascoli} Ascoli GA, Gasparini S, Medinilla V, Migliore M. 
   Local control of postinhibitory rebound spiking in CA1 pyramidal neuron dendrites. 
   {\it J Neurosci.}  {\bf 30} 6434 (2010).

\bibitem{2010Morse} Morse TM, Carnevale NT, Mutalik PG, Migliore M, Shepherd GM. 
   Abnormal Excitability of Oblique Dendrites Implicated in Early Alzheimer’s: A Computational Study. 
   {\it Front Neural Circuits.}  {\bf 4} 16 (2010).

\bibitem{1999Hoffman} Hoffman DA, Johnston D. 
   Neuromodulation of dendritic action potentials. 
   {\it J Neurophysiol.} {\bf 81} 408 (1999).

\bibitem{1999Magee} Magee JC. 
   Dendritic $I_{\rm h}$ normalizes temporal summation in hippocampal CA1 neurons. 
   {\it Nat. Neurosci.}  {\bf 2} 508 (1999).

\bibitem{2008Gunay} G\"unay C, Edgerton JR and Jaeger D.
   Channel Density Distributions Explain Spiking Variability in the Globus Pallidus: A Combined Physiology and Computer Simulation Database Approach
   {\em The Journal of Neuroscience} {\bf 28}(30), 7476 (2008).

\bibitem{1998Crook}  Crook SM, Ermentrout GB, Bower JM.
 Dendritic and Synaptic Effects in Systems of Coupled Cortical Oscillators. 
    {\em Journal of Computational Neuroscience} {\bf 5} 315 (1998)

\bibitem{1994Vreeswijk}  Vreeswijk C, Abbott LF, Ermentrout GB.
 When inhibition not excitation synchronizes neural firing. 
    {\em Journal of Computational Neuroscience} {\bf 1}(4) 313 (1994). doi:10.1007/bf00961879     

\bibitem{1996Buzsaki}  Wang X-J, Buzs\'aki G.
 Gamma Oscillation by Synaptic Inhibition in a Hippocampal Interneuronal Network Model. 
    {\em The Journal of Neuroscience} {\bf 16}(20) 6402 (1996).   

\bibitem{2019Prescott}  Lankarany M, Al-Basha D,  Ratt\'e S, and Prescott SA.
 Differentially synchronized spiking enables multiplexed neural coding. 
    {\em PNAS} {\bf 116}(20) 10097 (2019).   

\end{thebibliography}
\end{document}